\documentstyle[preprint,aps]{revtex}
\draft
\begin{document}
\title{Pairing of Fermions with Arbitrary Spin}
\author{$^{\ast +}$Tin-Lun Ho and $^{\ast}$Sungkit Yip}
\address{$^{\ast}$Physics Department, The Ohio State University,
Columbus, OH 43210\\$^{+}$ Institute for Theoretical Physics, University of
California at Santa Barbara, CA 93106}
\maketitle

\begin{abstract}
Motivated by the recent success of optical trapping of alkali Bose condensate, 
we have studied the superfluid state of optically trapped alkali fermions, 
which can have Cooper pairs with total spin $J\geq 2$. In this paper, we 
shall discuss the general structure of these large spin Cooper pairs and
their close relation with singlet Cooper pairs with non-zero 
orbital angular momentum. We also present 
the exact solution for the $J=2$ pairing which shows a surprising
change of ground state as the spin $f$ of the constituent fermion increases. 
\end{abstract}

The discovery of Bose-Einstein condensation\cite{BEC} in atomic gases has
stimulated many new research directions. Among these is the search of 
the superfluid phases of alkali fermions. This search has become even 
more exciting in view of the recent
success of confining Bose condensates in optical traps\cite{optical}.
Since optical traps are non-magnetic, the spin of the trapped atoms are
no longer frozen as they were in magnetic traps. This leads to a new class 
of superfluid phenomena. In the case of spin-1 Bose gas like $^{23}$Na 
and $^{87}$Rb, one of us\cite{Ho} has recently pointed out $^{23}$Na and 
$^{87}$Rb should have a non-magnetic and ferromagnetic {\em spinor} condensate 
respectively, according to the current estimates of their scattering 
lengths\cite{Greene}. Very recently, experiments at MIT\cite{Stenger} have 
verified the basically non-magnetic spinor nature of $^{23}$Na and 
found that its magnetic interaction is indeed antiferromagnetic\cite{Ho}. 

The physics of alkali fermions in optical traps is equally rich. The fact 
that all alkali fermions (except $^{6}$Li) have hyperfine spins (or simply 
``spins") $f>1/2$ in their lowest hyperfine manifold 
implies that their Cooper pairs can have total spins $J>1$.  
Fermions like $^{22}$Na and $^{134}$Cs which have $f=5/2$ and $7/2$ can have 
Cooper pairs with total spin as high as 4 and 6. From the example of 
superfluid $^{3}$He, one can be sure that the internal 
structure of these large spin Cooper pairs will generate a multitude of 
macroscopic quantum phenomena. The purpose of this paper is to point out the 
structure these large spin Cooper pairs, and a surprising change in behavior of 
a spin-$J$ Cooper pair as a function of fermion spin $f$. 

As a first step,  we shall focus on {\em homogenous} dilute Fermi gases in zero 
magnetic fields. It is important to understand 
the homogeneous situation before studying the trapped cases\cite{homo}.
Moreover, the physics 
of homogeneous systems are important in their own right. At first sight, the weak field 
limit seems difficult to achieve,  for even the Earth magnetic field amounts to 
$10^{-4}$K, enough to polarize the whole gas. 
Despite this ``strong" background field, which can be 
shielded off to a large extent, one can reduce it effectively  to the 
weak field limit by specifying the total spin $S$ of the system. 
Since the dynamics of these systems is spin conserving\cite{Ho}, a prepared 
spin $S$ out of equilibrium with an external field cannot relax 
to its equilibrium value.  The system therefore sees an effective field 
which would have been in equilibrium with the prepared $S$. 
By choosing $S$ appropriately, the effective field can be made much smaller 
than the external one. This method has very recently been used by Ketterle's 
group to study the spinor nature of the $^{23}$Na condensate\cite{Stenger}.          

Of course, for a pairing state to be observable, its pairing interaction 
has to be sufficiently negative to produce an observable $T_{c}$. 
While the scattering lengths of some alkali fermions have been calculated,
they remain unknown for many alkalis.  (See later discussion). 
In view of the lack of information, we have performed a general 
study of the large spin Cooper pairs. 
In particular, we shall discuss the $J=2$ pairing in detail. 
This is the simplest among all large spin pairing which also has an exact 
solution. The phenomena contained in this case reveal the rich physics 
installed, which turns out to be remarkable indeed.  
For simplicity, we shall call the S-wave spin-$J$ Cooper pairs (made up of
two spin $f$ fermions) ``spin" Cooper pairs, and  
singlet Cooper pairs with 
{\em orbital} angular momentum $J$ (made up of two spin-1/2 fermions) 
``orbital" Cooper pairs.  Let us first 
summarize our findings : 

\noindent {\bf (A)} The  structure of ``spin" Cooper pairs is analogous to 
that of ``orbital" Cooper pairs with the same angular momentum. 
This allows one to obtain information of the former from the latter, 
for which an exact solution already exists for $J=2$. 

\noindent {\bf (B)} The structure of a spin-$J$ Cooper 
changes as the spin $f$ of the constituent fermions increases 
beyond a critical value. 
For  Cooper pairs with spin $J=2$, they are ``ferromagnetic" (or ``axial")
 if $f\geq 7/2$,  
but non-magnetic (or ``real") if
$f\leq 5/2$. This change of character as a function of $f$ is a result 
of maximizing the 
phase space for pairing and is {\em independent} of interaction 
parameters, as long as 
they favor $J=2$ pairing. 

{\em Free energy :}  The low energy effective Hamiltonian of a 
spin-$f$ dilute Fermi gas with s-wave interactions has been derived in 
ref.\cite{Ho}. It is rotationally invariant in spin space, and is 
of the form 
$H-\mu N$$=\int{\rm d}{\bf x} \psi^{+}_{\alpha}({\bf x}) 
{\cal H}^{o}_{\alpha \beta}({\bf x}) 
\psi_{\beta}({\bf x})$ 
 $+\frac{1}{2}\int{\rm d}{\bf x}\psi^{+}_{\alpha}({\bf x}) 
  \psi^{+}_{\beta}({\bf x}) 
\Gamma_{\alpha\beta; \mu\nu}
\psi_{\mu}({\bf x}) \psi_{\nu}({\bf x})$, 
${\cal H}^{o}_{\alpha \beta}({\bf x})$$= -\frac{\hbar^{2}}{2M}{\bf \nabla}^{2}
\delta_{\alpha\beta} - \gamma{\bf B}\cdot{\bf F}_{\alpha\beta}$, 
\begin{equation}
\Gamma_{\alpha\beta; \mu\nu} = \sum_{F=0}^{2f-1} g_{F} \sum_{m=-F}^{F}
<ff \alpha\beta|ff;Fm><ff;Fm|ff \mu\nu>
\label{Gamma} \end{equation}
where $M$ is the mass of the fermion, 
$<ff;Fm|ff \mu\nu>$ is the Clebsch-Gordan coefficient for forming 
a total spin $F$ from two spin-$f$ particles,  $g_{F}=4\pi\hbar^{2}a_{F}/M$,  
and $a_{F}$ is the s-wave scattering length of two spin-$f$ fermions
in the scattering channel with total spin $F$. Because of antisymmetry of the
fermions, only even $F$'s appear in eq.(\ref{Gamma}). 

The order parameter of an S-wave superfluids is 
$\Psi_{\alpha\beta}({\bf x}) = <\psi_{\alpha}({\bf x})
\psi_{\beta}({\bf x})>$, which is 
a $(2f+1)\times(2f+1)$ antisymmetric matrix in spin space. 
For homogeneous systems, $\Psi_{\alpha\beta}$ is 
independent of ${\bf x}$. It is convenient to define
the gap function 
\begin{equation}
\Delta_{\alpha\beta} = \Gamma_{\alpha\beta; \mu\nu}\Psi_{\mu\nu}. 
\end{equation}
The free energy according BCS theory is 
\begin{equation}
{\cal F} = \frac{1}{2}{\rm Tr} \Delta^{+}\Gamma^{-1}\Delta 
-\frac{k_{B}T}{2} \sum_{{\bf k}\omega_{n}} \sum_{\ell=1}^{\infty}
\frac{1}{\ell}{\rm Tr}\left[\Delta\tilde{G}({\bf k}\omega_{n})\Delta^{+}G({\bf
k}\omega_{n})\right]^{\ell}
\label{F} \end{equation}
where $\Gamma^{-1}$ is given by eq.(\ref{Gamma}) with $g_{F}$ replaced by
$g_{F}^{-1}$, $\omega_{n}=(2n+1)\pi k_{B}T$ are the Matsubara frequencies, 
$G_{\alpha\beta}({\bf k}, \omega_{n})$ and 
$\tilde{G}_{\alpha\beta}({\bf k}, \omega_{n})$ are normal Greens functions 
satisfying matrix equation
$\left(i\omega_{n} - {\cal H}_{o}({\bf k})\right)G({\bf k}\omega_{n})
= 1$, 
$\left(i\omega_{n} + {\cal H}^{T}_{o}({\bf k})\right)
\tilde{G}({\bf k}\omega_{n}) =1$.  

{\em General structure of Cooper pairs with spin angular momentum $J$ :}
Under a spin rotation 
$U={\rm exp}\left( -i\vec{\theta}\cdot {\bf F}\right)$, 
$\psi_{\alpha}$$\rightarrow$$(U\psi)_{\alpha}$. This implies 
$\Psi \rightarrow \Psi'=U\Delta U^{T}$, and hence 
$\Delta \rightarrow \Delta'=U\Delta U^{T}$. For gap functions that transform
like an angular momentum state $|Jm>$, they must satisfy
\begin{equation}
\left[U \Delta^{(J)}_{m} U^{T}\right]_{\alpha\beta}
= D^{(J)}_{mm'}(\vec{\theta}) \left[  \Delta^{(J)}_{m'} \right]_{\alpha\beta}
\label{trans} \end{equation}
It is easy to see that the solution of eq.
(\ref{trans}) is 
$(\Delta^{(J)}_{m})_{\alpha\beta} \propto <ff\alpha\beta|ff;Jm>$. The general 
structure of the spin-$J$ gap function is therefore 
\begin{equation}
(\Delta^{(J)})_{\alpha\beta} \propto \sum_{m=-J}^{J} c_{m} 
<ff\alpha\beta|ff;Jm>, 
\,\,\,\,\,\,\, {\rm or} \,\,\,\,\,
|\Delta^{(J)}> \propto \sum_{m=-J}^{J} c_{m}|Jm>,
\label{expand} \end{equation}
where the second expression in eq.(\ref{expand}) is simply the first written 
in abstract form. 

To find the $\Delta^{(J)}$ that minimizes the energy, and to 
illustrate the relation of ``spin" and ``orbital" Cooper pairs, it is 
useful to consider a different representation of $\Delta^{(J)}$. 
First, we note that the singlet state $(\Delta^{(0)})_{\alpha\beta}\propto
 \eta \equiv <ff;00|ff\alpha\beta>\sqrt{2f+1}$
satisfies $U\eta U^{T}=\eta$, and has the properties $\eta^{+}\eta=1$, 
$U^{+} \eta = \eta U^{T}$, and $F_{i}\eta = - \eta F_{i}^{T}$. 
Defining $\Delta \equiv \Xi\eta$, Eq.(\ref{trans}) then becomes  
$U\Xi^{(J)}_{m} U^{+}= D^{(J)}_{mm'}\Xi^{(J)}_{m'}$, which has the solution
$\left[\Xi^{(J)}_{m}\right]_{\alpha\beta} \propto \left[Y_{Jm}({\bf F})
\right]_{\alpha\beta}$, where $Y_{Jm}({\bf F})$ is a matrix obtained 
by first writing the spherical harmonic $k^{J}Y^{(J)}_{m}(\hat{\bf k})$ in a
symmetric rectangular form, 
and then by replacing $k_{i}$ by the matrix $F_{i}$\cite{Baym}. 
For example, since $k^{2}Y_{21}(\hat{\bf k}) \propto  
k_{z}(k_{x} +  ik_{y})$, we have 
$Y_{21}(\hat{\bf F}) \propto F_{z}(F_{x} +  iF_{y}) + (F_{x} +  iF_{y})F_{z}$. 
The general form of the order parameter within the angular momentum $J$ subspace
is then 
\begin{equation}\Delta_{\alpha\beta}^{(J)} = \sum_{m=-J}^{J}  c_{m}\left[ 
Y_{Jm}({\bf F}) \eta \right]_{\alpha\beta}.   \label{sphere}
\end{equation}
Using Wigner-Eckart theorem, it is easily seen that the two representation 
eq.(\ref{expand}) and eq.(\ref{sphere}) are identical.

Next, we note that $r^{J}Y_{Jm}(\hat{\bf r})$ is a homogenous 
polynomial of ${\bf r}$ satisfying Laplace's 
equation. It can therefore be written as $r^{J}Y_{Jm}(\hat{\bf r})
= A_{i_{1}i_{2} ... i_{J}}r_{i_{1}}r_{i_{2}}...r_{i_{J}}$, where 
$A_{i_{1}i_{2} ... i_{J}}$ is {\em symmetric in all its indices and  
vanishes whenever any two indices contract}. We can then write $\Delta^{(J)}$ as 
\begin{equation}
\Delta^{(J)} = \sum_{i_{1} ... i_{J}} 
A_{i_{1}i_{2} ... i_{J}}F_{i_{1}}F_{i_{2}}...F_{i_{J}}\eta . 
\label{AF} \end{equation}
It is also useful to compare the {\em spin} structure of $\Delta^{(J)}$ in 
eq.(\ref{AF})  with the {\em orbital} structure of the singlet
Cooper pairs of spin-1/2 fermions. The order parameter of the latter is 
$\Delta({\bf k}) =<c_{\uparrow}({\bf k})c_{\downarrow}(-{\bf k})>$, where 
$c^{+}_{\uparrow}({\bf k})$ creates a spin +1/2 fermion with momentum ${\bf k}$
at the Fermi surface. For pairing with {\em even} orbital angular momentum $J$, 
$\Delta^{(J)}({\bf k}) = \sum_{m} c_{m}Y_{Jm}(\hat{\bf k})$, or
\begin{equation}
\Delta^{(J)}({\bf k}) = \sum_{i_{1} ... i_{J}}  A_{i_{1}i_{2} ...
i_{J}}k_{i_{1}}k_{i_{2}}...k_{i_{J}}. 
\label{Lwave} \end{equation}
Comparing eq.(\ref{AF}) and eq.(\ref{Lwave}), one finds that they 
are almost identical except that the  $F_{i}$'s 
are non-commuting matrices whereas the $k_{i}$s are c-numbers.
On the other hand, this means that these two structures approach  each other 
as $f$ increases, since the 
spin operator ${\bf F}$ behaves more like a classical vector. 

{\em The general scheme for determining $\Delta^{J}$ and the 
$J=2$ pairing  :}  
At temperature is lowered,  
superfluid condensation first takes place at the (even) angular momentum 
channel $J$ with largest negative coupling constant $g_{J}$.  
The free energy eq.(\ref{F}) to the quartic order in $\Delta^{(J)}$ is
\begin{equation} 
{\cal F} =  -\frac{1}{2}\alpha {\rm Tr} \Delta^{(J)+}\Delta^{(J)}  
 + \frac{1}{4}\beta {\rm Tr}(\Delta^{(J)} \Delta^{(J)+})^{2}  
\label{FJ} \end{equation}
where $\alpha = N(0){\rm ln}(T_{c}/T)$, $T_{c} =
1.14\epsilon_{F}e^{-1/(\vert g_{J}\vert N(0))}$$=1.14\epsilon_{F}
e^{-\pi/(2k_{F}\vert a_{J} \vert )}$, $N(0)$ 
is the density of state at the
fermi surface per spin, $\beta = 7\zeta(3)/(8\pi^{2}T_{c}^{2})$, and 
$\epsilon_{F}$ and $k_{F}$ are the Fermi energy and momentum.  
To determine $\Delta^{(J)}$, we substitute eq.(\ref{AF}) into eq.(\ref{FJ}) and
find the matrix $A$ that minimizes the energy. 
In the following, we shall present the exact solution for
for S-wave $J=2$ Cooper pairs formed by spin-$f$ fermions. 
The solutions of $J>2$ 
Cooper pairs will be studied elsewhere for they require much lengthier 
calculations than the $J=2$ case, which is already lengthy. 
Our method, however, applies to all $J\geq 2$ pairs. 

>From eq.(\ref{FJ}), one can see that ${\cal F}$ is of the form 
\begin{equation}
{\cal F} =  - \frac{\alpha}{2} 
A_{ij}A^{\ast}_{pq} {\rm Tr}(F_{i}F_{j}F_{p}F_{q}) 
+ \frac{\beta}{4}
A_{ij}A^{\ast}_{k\ell}A_{pq}
A^{\ast}_{st}
{\rm Tr}(F_{i}F_{j}F_{k}F_{\ell}F_{p}F_{q}F_{s}F_{t}) ,  
\label{quartic} \end{equation}
After evaluating the traces (see Appendix), we find 
\begin{equation}
{\cal F} =  - \tilde{\alpha} {\rm Tr}AA^{+} + 
\beta_{1}\left|{\rm Tr}A^{2}\right|^{2} +
\beta_{2}\left({\rm Tr}A^{\ast}A\right)^{2} + 
\beta_{3}{\rm Tr}\left(A^{2}A^{\ast 2}\right)
\label{dwave} \end{equation}
\begin{equation}
\beta_{1} = \frac{\beta}{4} 
\left[ -\frac{29}{70}I_{2} + \frac{121}{60}I_{4} - 
\frac{22}{15}I_{6} + \frac{4}{35}I_{8}
  \right]
\label{beta1}\end{equation}
\begin{equation}
\beta_{2} = \frac{\beta}{4} 
\left[ -\frac{2}{70}I_{2} + \frac{1}{30}I_{4} + 
\frac{4}{15}I_{6} + \frac{8}{35}I_{8}  \right]
\label{beta2} \end{equation}
\begin{equation}
\beta_{3} = \frac{\beta}{4} 
\left[ \frac{3}{5}I_{2} - \frac{8}{3}I_{4} + 
\frac{16}{15}I_{6}  \right] \label{beta3} 
\end{equation}
where $I_{n} \equiv \sum_{m=-f}^{f} m^{n}$, and
 $\tilde{\alpha}= {\alpha \over 12} [ 4 I_4 - I_2 ] $.

Eq.(\ref{dwave}) is identical 
to the free energy of a general {\em d-wave} singlet superfluids
The minimization problem of eq.(\ref{dwave}) was
solved by Mermin\cite{Mermin}. Only three equilibrium 
phases are possible
\cite{sym}:

\noindent ({\bf I}) ``Axial" state : 
When  $\beta_{3}>-\beta_{1} + |\beta_{1}|$,, 
$\Delta \propto Y_{22}({\bf F})\eta$,

\noindent ({\bf II}) ``Cyclic" state : 
When $0>\beta_{3}>-6\beta_{1}$, $\Delta \propto 
(F_{x}^{2}+ e^{2\pi i/3}F_{y}^{2} + e^{4\pi i/3}F_{z}^{2})\eta$, 

\noindent ({\bf III}) ``Real" state : 
When $\beta_{3}< - 4\beta_{1} -2|\beta_{1}|$, 
$\Delta \propto [\zeta_{1} Y_{20}({\bf F}) + \zeta_{2}(Y_{22}({\bf F}) 
+ Y_{2,-2}({\bf F}))]\eta$, where $\zeta_{1}$ and $\zeta_{2}$ are real.

\noindent The portion of
the phase diagram  in $\beta_{1}-\beta_{3}$ space relevant for our
discussion is shown in fig.1. Using eq.(\ref{beta1}) 
to eq.(\ref{beta3}), we note
that $(\beta_{1}, \beta_{3})$ is in region {\bf III} 
for $f=\frac{3}{2}$ and $\frac{5}{2}$, 
and in   region {\bf I} for $f \geq \frac{7}{2}$. The 
superfluid is therefore a 
``real" state for $f=\frac{3}{2}$ and $\frac{5}{2}$, but changes to the 
``axial" state 
for $f \geq \frac{7}{2}$. 

This change of pairing behavior can be understood as follows. 
As mentioned before, as 
$f\rightarrow \infty$, the order parameters 
in eq.(\ref{AF}) and eq.(\ref{Lwave}) 
become identical,  and that the
energy eq.(\ref{FJ}) becomes that the weak coupling d-wave superfluid, 
which has an optimum order parameter 
$Y_{22}(\hat{\bf k})$\cite{Mermin}. (This state has ``more pairing" 
 $Y_{20}$ and $Y_{2\pm 1}$ in the sense that its absolute square 
only has point nodes
whereas both $|Y_{20}(\hat{\bf k})|^{2}$ 
and $|Y_{2,\pm 1}(\hat{\bf k})|^{2}$ have line nodes). 
On the other hand, in the most quantum case $f=\frac{3}{2}$, there are four
degenerate Fermi surfaces, labelled by $m_{z}=\pm \frac{3}{2}, \pm\frac{1}{2}$. 
The structure of the  axial and the real state are given by
$|\Delta_{axial}> = |2; 2> = \frac{1}{\sqrt{2}}\left( 
|\frac{3}{2}, \frac{1}{2}>-|\frac{3}{2}, \frac{1}{2}>\right)$, 
$|\Delta_{real}> = \zeta_{1}|2; 0> + \zeta_{2}\left( 
|2;2>+ |2,-2>\right)$$={\zeta_{1} \over 2} \left[ 
\left(|\frac{3}{2}, \frac{-3}{2}>
- |\frac{-3}{2}, \frac{3}{2}> \right) +  \left(|\frac{1}{2}, \frac{-1}{2}>
- |\frac{-1}{2}, \frac{1}{2}> \right) \right] $
$+{\zeta_{2} \over 2} \left[ 
\left( |\frac{3}{2}, \frac{1}{2}> -  |\frac{1}{2}, \frac{3}{2}>\right)
+ \left( |\frac{-1}{2}, \frac{-3}{2}> -  
|\frac{-3}{2},\frac{-1}{2}>\right)\right]$, 
where the state vectors with integer entries such as $|2;0>$ means 
$|\frac{3}{2}\frac{3}{2}; J=2,
m=0>$, those with half integer entries such as 
$|\frac{1}{2},\frac{-1}{2}>$ means
 $|f=\frac{3}{2}, m=\frac{1}{2}>|f=\frac{3}{2};m= \frac{-1}{2}>$. 
One can see that the only two Fermi surfaces ($m=\frac{3}{2}$ and
$\frac{1}{2}$) are involved in the pairing in axial state, 
whereas all four Fermi surfaces are involved in the pairing of the real state. 
Since the real state maximizes the amount of pairing, it is 
favored in this
extreme quantum case. As $f$ increases, the number of Fermi surfaces 
appearing in the 
axial state (i.e the spin state $|J=2, m=2>)$ quickly increases. 
By the time when $f$ reaches $\frac{7}{2}$, the real state 
no longer has the advantage of
involving most Fermi surfaces, and the system switches to the axial state, where the 
spin operator ${\bf F}$ begins to resemble a classical vector. 
We have thus established Statements ({\bf A}) and ({\bf B}).

{\em Observability :}  The long lived alkali fermions which have 
$f> \frac{1}{2}$ in their lowest hyperfine manifold are 
$^{22}$Na, $^{40}$K, $^{86}$Rb, $^{132}$Cs, $^{134}$Cs, and $^{136}$Cs, 
which have $(f=5/2, 9/2, 5/2, 3/2, 7/2, 9/2)$ and lifetimes
(2.5 yrs, $10^{9}$ yrs, 18 days, 6 days, 2 yrs, 13 days) 
respectively\cite{Table}.  
According to the recent calculation of
Greene, Burke, and Bohn\cite{Greene}, the scattering lengths of $^{40}$K are 
positive, hence unfavorable for pairing. At present, there is no information
about the scattering lengths of the Cs fermions.  On the other hand, 
$a_{4}= -65 (+40, -120) a_{B}$,
$a_{2}= -130 (+40, -70) a_{B}$, 
$a_{0}= -145 (+40, -65) a_{B}$ for $^{86}$Rb; and 
$a_{4}= -108 (+27, -40) a_{B}$,
$a_{2}= -115 (+32, -50) a_{B}$, 
$a_{0}= -117 (+34, -55) a_{B}$ for $^{22}$Na , 
where $a_{B}$ is the Bohr radius and the numbers
in the bracket  are error bars.  

To estimate $T_c$. we use the value of $k_{F}$ and
$\epsilon_{F}$ {\em at the center of the trap}. 
For an anisotropic trap with frequencies $\omega_{\perp}$ and $\omega_{z}$ 
in the $xy$-plane and along $z$, it is easy to show that
$k_{F}a_{2}=({R \over a_{\perp} })({a_{2} \over a_{\perp}})$, 
$\epsilon_F = \frac{1}{2}\hbar \omega_{\perp}
({R \over a_{\perp} })^2$, where $a_{\perp}=\sqrt{\hbar/M\omega_{\perp}}$, 
 where $R$ 
is the radius of the cloud in the $xy$-plane 
related to the total number
of particles $N$ as
${R \over a_{\perp}} = \left({ 48 N  \lambda \over  (2f + 1)}\right)^{1/6}$, 
with $\lambda \equiv \omega_{z}/\omega_{\perp}$.
For an isotropic  trap ($\lambda=1$)
with $\omega_{\perp}/2\pi =2000$Hz,
the expression 
$T_{c}(J=2) = 1.14\epsilon_{F}e^{-\pi/2 k_{F} \vert a_{2} \vert }$ gives
$T_{c}(J=2) \sim 1.9 \times
10^{-8} K$ for $^{22}$Na with $N=4\times 10^{6}$ atoms  and  
$T_{c}(J=2) \sim 2.3 \times
10^{-7} K$ for $^{86}$Rb with $N=10^{6}$ atoms. 
Since the lowest
temperature reached in current BEC experiments is $10^{-9}$K, these transition 
temperatures of fermions (which can be made higher by increasing the trap
frequency or the anisotropy $\lambda$) appear to be feasible. 
Since $g_{0}$ is most negative, singlet instead of $J=2$ 
pairing will first occur in zero field. 
This, however, does not mean that all higher spin pairing states are 
non-observable. The singlet
spin states can be efficiently suppressed in a magnetic field 
(obtained  by specifying the total
spin of the system as mentioned in the introduction), 
thereby revealing all other higher spin pairing states\cite{mag}.
For length reasons, the effects of spin constraints 
 will be discussed elsewhere. 

{\em Final Remarks}: We have shown that the superfluid phenomena of alkali 
fermions become amazingly rich once the spin degrees of freedom are released. 
Should the current efforts of cooling alkali fermions to degenerate limit 
be successful, transferring the degenerate gas into an optical 
trap\cite{optical} will help one to uncover the superfluid 
phases discussed here.  Since $^{132}$Cs, 
$^{134}$Cs, and $^{136}$Cs have $f=3/2, 7/2$, and $9/2$ in their lowest 
hyperfine multiplet respectively, if their scattering lengths turned out to 
be negative, our result predicts that like $^{22}$Na and $^{86}$Rb, the ground 
state of $^{132}$Cs will a ``real" state, whereas 
$^{134}$Cs, and $^{136}$Cs will be an ``axial" state.

{\em Appendix: Evaluation of the quartic term in eq.(\ref{quartic}) :} Denoting 
${\cal I}_{4}=A_{ij}A^{\ast}_{k\ell}A_{pq}
A^{\ast}_{st} {\rm Tr}(F_{i}F_{j}F_{k}F_{\ell}F_{p}F_{q}F_{s}F_{t})$,
we note that ${\cal I}_{4} = {\cal D} \left[ {\rm Tr}U\right]_{o}$, 
where  ${\cal D}\equiv A_{ij}A^{\ast}_{k\ell}A_{pq}A^{\ast}_{st} 
\frac{\partial^{2}}{\partial a_{i}\partial a_{j}}
\frac{\partial^{2}}{\partial b_{k}\partial b_{\ell}}
\frac{\partial^{2}}{\partial c_{p}\partial c_{q}}
\frac{\partial^{2}}{\partial d_{s}\partial d_{t}}$, 
 $U\equiv e^{-i{\bf a}\cdot{\bf F}}e^{-i{\bf b}\cdot{\bf F}}
e^{-i{\bf c}\cdot{\bf F}}e^{-i{\bf d}\cdot{\bf F}}\equiv 
e^{-i\vec{\theta}\cdot {\bf F}}$, and 
the subscript $``o"$ means ${\bf a}={\bf b}={\bf c}={\bf d}=0$.
Expanding $U$ in
powers of $\vec{\theta}$, it is easy to see that ${\cal I}_{4} = \sum_{n=1}^{4}
\frac{(-1)^{n}}{(2n)!} I_{2n} \left({\cal D}\theta^{2n}\right)_{o}$, where 
$I_{2n} = \sum_{m=-f}^{f}m^{2n}$. Next, we note that 
the relation between $\theta$ and ${\bf a, ..,  d}$
is independent of $f$. For spin 1/2 systems, the
quantity $Q$$\equiv$${\rm Tr}$$[e^{-i{\bf a}\cdot\vec{\sigma}/2}$
$e^{-i{\bf b}\cdot\vec{\sigma}/2}$
$e^{-i{\bf c}\cdot\vec{\sigma}/2}$$e^{-i{\bf
d}\cdot\vec{\sigma}/2}]$$/{\rm Tr}(1)$
$=<e^{-i\vec{\theta}\cdot\vec{\sigma}/2}>$ 
can be written as $\xi=Q-1=\sum_{n=1, 2, ..}\frac{(-1)^{n}}{2^{2n}(2n)!}
\theta^{2n}$.  Inverting this relation, we obtained $\theta^{2}$ as a power 
series of $\xi$, or $Q$. 
From this expression, we can calculate 
$({\cal D}\theta^{2n})_{o}$ for $n=2$ to $4$ by calculating
${\cal D}Q^{p}$ for $p=1,2,3,4$.  The latter can be easily calculated because
they involve only spin 1/2 quantities. Evaluating  ${\cal I}_{4}$ this way 
gives eq.(\ref{beta1}) to
eq.(\ref{beta3}).

Acknowledgement : TLH would like to thank Jim Burke for the estimates of the 
scattering lengths. 
This work is supported by a Grant
from NASA NAG8-1441, and NSF Grants DMR-9705295 and  DMR-9807284.

\vspace{0.2in}

\noindent {\bf Figure caption}:
Distribution of stable phases in the $(\beta_{1}/\beta_{2}, 
\beta_{3}/\beta_{2})$ space for $J=2$ pairing\cite{Mermin}.
Regions ${\bf I, II, III} $ are the stable regions of the ``axial", 
``cyclic", and ``real" states resp. 
${\bf U}$ corresponds to an unstable region.


\begin{thebibliography}{99}
\bibitem{BEC} M.H. Anderson, J.R. Ensher, M.R. Mathews, C.E.
Weiman, and  E.A. Cornell, Science {\bf 269}, 198 (1995).
K. B. Davis, M.-O. Mewes, M.R. Andrews, N.J. van Druten, D.S.
Durfee, D.M. Kurn,  and W. Ketterle,  Phys. Rev. Lett. {\bf 75}, 3969 (1995).
\bibitem{optical} D. Stamper-Kurn. M.R. Andrews, A.P. Chikkatur, S. Inouye, H.J.
Miesner, J. Stenger, and W. Ketterle. Phys.Rev. Lett.{\bf 80}, 2027 (1998). 
\bibitem{Ho} T.L. Ho, Phys. Rev. Lett. {\bf 81}, 742, (1998).
\bibitem{Greene} J. Burke, C. Greene, and J. Bohn, to be published. 
\bibitem{Stenger} J. Stenger, et.al. {\em Spin domains in ground state spinor
Bose-Einstein condensates}, preprint. 
\bibitem{homo}  For not too small number of particles,
the local fermi wavelength will be much less than the size of
the cloud almost everywhere.
  One can therefore think of the order parameter locally. 
The proximity effect will however lower the $T_c$ somewhat
from the estimate discussed below.
\bibitem{Baym} G. Baym, 
Ch.17, {\em Lectures on Quantum Mechanics}, Benjamin/Cummings, 1969.
\bibitem{Mermin} N.D. Mermin, Phys. Rev. A {\bf 9}, 868 (1974).
\bibitem{sym}  The phases listed consist of
degenerate manifolds. A general state is obtained from
those listed by rotation and gauge transformations.
\bibitem{Table} {\em Nuclear Wallet Cards}, 5th ed., edited by 
Jagdish K. Tuli, Brookhaven National Laboratory.
\bibitem{mag} More precisely, the $J=0$ transition will be more strongly
suppressed than the axial state $\Delta \propto Y_{22}({\bf F})\eta$
or the real state $\Delta \propto [Y_{22}({\bf F}) + Y_{2-2}({\bf F}) ]\eta$.
\end{thebibliography}
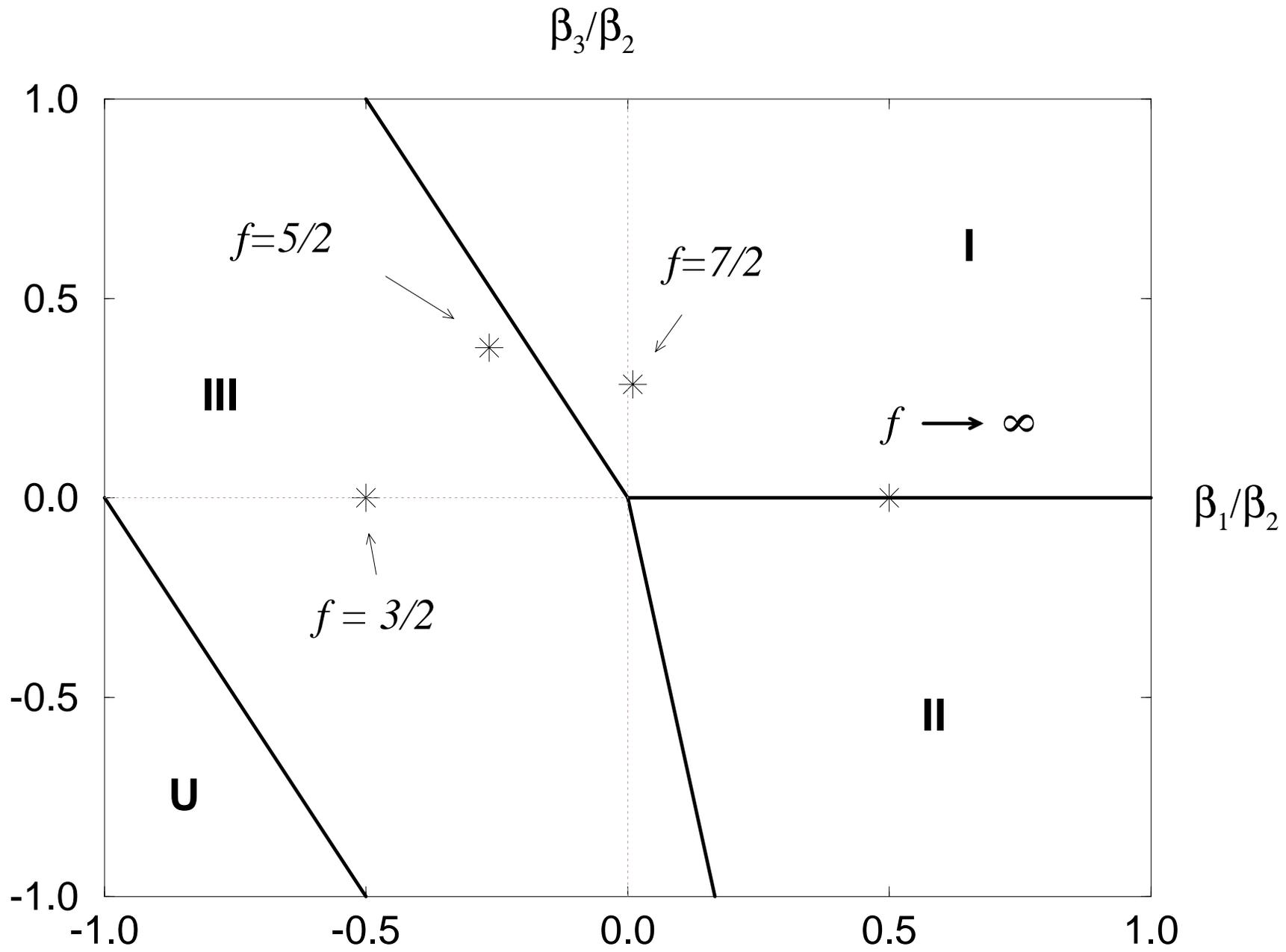
\end{document}